\documentclass{llncs}
\usepackage{amssymb}
\usepackage{amsmath}
\usepackage{multirow}
\usepackage[inline]{enumitem}

\usepackage{graphicx}

\usepackage{colortbl}
\usepackage{ctable}
\usepackage{soul}
\usepackage{footmisc}
\usepackage{siunitx,booktabs,etoolbox}
\usepackage[misc]{ifsym}
\newcommand{\mat}[1]{{\boldsymbol{{#1}}}} % matrix

\providecommand{\mx}{\mat{x}}

\providecommand{\my}{\mat{y}}
\providecommand{\mz}{\mat{z}}

\usepackage{mathtools}
\usepackage{ifsym}
\begin{document}

%______________________________________________________________________________________________________________________

\title{Data Efficient Unsupervised Domain Adaptation for Cross-Modality Image Segmentation}

%\titlerunning{Data Efficient Unsupervised Domain Adaptation for Image Segmentation}
%\authorrunning{C. Ouyang et al.   }
% \author{*** *** *** *** *** *** *** ***}
\author{Cheng Ouyang\inst{1}$^{\textrm{(\Letter)}}$ 
	\and Konstantinos Kamnitsas\inst{1}
	\and Carlo Biffi\inst{1}
	\and Jinming Duan\inst{1,2}
	\and Daniel Rueckert\inst{1}}
	%index{Ouyang,Cheng}
	%index{Kamnitsas,Konstantinos}
	%index{Biffi,Carlo}
	%index{Duan,Jinming}
	%index{Rueckert,Daniel}
\institute{Biomedical Image Analysis Group, Imperial College London, UK \\
\and School of Computer Science, University of Birmingham, UK\\
\email{c.ouyang@imperial.ac.uk}}
\maketitle 
\begin{abstract}
Deep learning models trained on medical images from a source domain ($\mathit{e.g.}$ imaging modality) often fail when deployed on images from a different target domain, despite imaging common anatomical structures. Deep unsupervised domain adaptation (UDA) aims to improve the performance of a deep neural network model on a target domain, using solely unlabelled target domain data and labelled source domain data. However, current state-of-the-art methods exhibit reduced performance when target data is scarce. In this work, we introduce a new data efficient UDA method for multi-domain medical image segmentation. The proposed method combines a novel VAE-based feature prior matching, which is data-efficient, and domain adversarial training to learn a shared domain-invariant latent space which is exploited during segmentation. Our method is evaluated on a public multi-modality cardiac image segmentation dataset by adapting from the labelled source domain (3D MRI) to the unlabelled target domain (3D CT). We show that by using only one single unlabelled 3D CT scan, the proposed architecture outperforms the state-of-the-art in the same setting. Finally, we perform ablation studies on prior matching and domain adversarial training to shed light on the theoretical grounding of the proposed method.
\end{abstract}
\section{Introduction}
Ideally, deep learning models deployed in medical imaging applications should be invariant to image appearance shifts caused by reasons such as different imaging modalities, scanning protocols or demographic properties. 
Unfortunately, in reality, deep learning usually suffers from the domain shift problem \cite{ben2010theory}. Given two different input domains with data $X$ and distribution $P(X)$, $\mathcal{D}_S = \{X_S, P(X_S)\}$, $\mathcal{D}_T = \{X_T, P(X_T)\}$ and a shared label space $\mathcal{Y} = \{Y \}$, a predictive model $f(\cdot)$ which approximates $P(Y | X)$ trained on the source domain $\mathcal{D}_S$ is likely to underperform on the target domain $\mathcal{D}_T$ when the distribution of data in $\mathcal{D}_T$ is different ($\mathit{e.g.}$ image appearance differences as described above). In this case, to transfer the source model to the target domain, target data and corresponding labels $\{(\mx_T, \my_T)\}$ are necessary for supervised fine-tuning-based transfer learning. In many settings though, such as medical imaging applications, manual labelling for target images is usually prohibitively expensive or impractical. This motivates \textit{unsupervised domain adaptation} (UDA), a methodology that seeks to learn a model that performs well in a target domain using solely \textbf{unlabelled} target domain data $\{ \mx_T \}$, besides any labelled data available in source domain. 

UDA usually assumes an underlying domain-invariant feature space $\mathcal{Z}$, which can be projected from $\mathcal{D}_S$ and $\mathcal{D}_T$ and can be utilized for a specific task. The most popular way to perform UDA is therefore learning mappings $\{ h_S(\cdot)$, $h_T(\cdot) \}$ from $\mathcal{D}_S$ and $\mathcal{D}_T$ to $\mathcal{Z}$ by matching their distributions in $\mathcal{Z}$ under certain distance metrics ($\mathit{e.g.}$ Jensen-Shannon distance).
Using this framework, \cite{tzeng2014deep} proposes to minimize the Maximum Mean Discrepancy (MMD) between source and target feature representations. With recent significant advancement of generative adversarial networks (GAN), distances between source and target domain can be estimated and minimized with domain adversarial training \cite{ganin2016domain,tzeng2017adversarial}, where the discriminator differentiates the domain of its input, while the generator generates domain-invariant representations to confuse the discriminator. Inspired by the work in \cite{zhu2017unpaired}, \cite{hoffman2017cycada} further promotes the performance by retaining semantic information of feature maps during domain transfers, by enforcing cycle-consistencies. 

\noindent{\bf Related work:} In medical image analysis, recent related works are mainly based on domain adversarial training. They are designed to mitigate domain gaps including modalities \cite{cai2019towards,chen2019synergistic,dou2018pnp,zhang2018task}, scanning protocols \cite{kamnitsas2017unsupervised}, and cross-center differences \cite{dong2018unsupervised}. The most recent state-of-the-art methods is \textit{SIFA} \cite{chen2019synergistic}, which is designed for medical image segmentation and is reported to outperform peer methods designed for natural images. It uses cycle-consistency as in \cite{hoffman2017cycada} and further employs a synergy of image-level and feature-level domain adversarial training. 
However, these methods suffer from an idealized assumption that abundant target data $\{\mx_B \}$ is always available, which is not always realistic in clinic practice. %data-collection is time-consuming and relies on a number of uncontrollable factors. 
Current pure data-driven, adversarial UDA is sub-optimal in such low-resource setting as data-driven GANs become inaccurate with small amounts of samples. 

\noindent\textbf{Contributions:} In this work, we for the first time investigate the challenging problem of \textit{UDA with scarce target data} in medical image segmentation. We propose a novel data-efficient UDA method for it. We focus on mitigating domain gaps manifested by differences in image appearance, of which cross-modality difference is a typical example. To compensate for the drawback of domain adversarial training given only a small number of target samples, we propose to introduce prior regularization on a shared feature space of the source and target domain images where segmentation is operated on. By independently enforcing the prior distributions for the source features and target features to be close to a fixed prior distribution (in our case, $\mathcal{N}(0,I)$), the prior regularization serves as an additional constraint for distribution matching. This constraint is in particular data-efficient, since KL-divergences from source or target feature distributions to $\mathcal{N}(0,I)$ can be estimated analytically. To easily obtain and to fully exploit this prior matching effect, we propose to 
\begin{enumerate*}[label=(\roman*),before=\unskip{ }, itemjoin={{, }}, itemjoin*={{, and }}]
  \item combine variational autoencoder (VAE), whose prior distribution of latent space can be analytically regularized, with domain adversarial training
  \item to directly operate image segmentation in this VAE latent space
 \end{enumerate*}, for UDA in cross-modality medical image segmentation.
%______________________________________________________________________________________________________________________
% Section 2: Method
\section{Method}
\begin{figure}[tb]
\centering
\scalebox{0.91}{\includegraphics[width=1\textwidth]{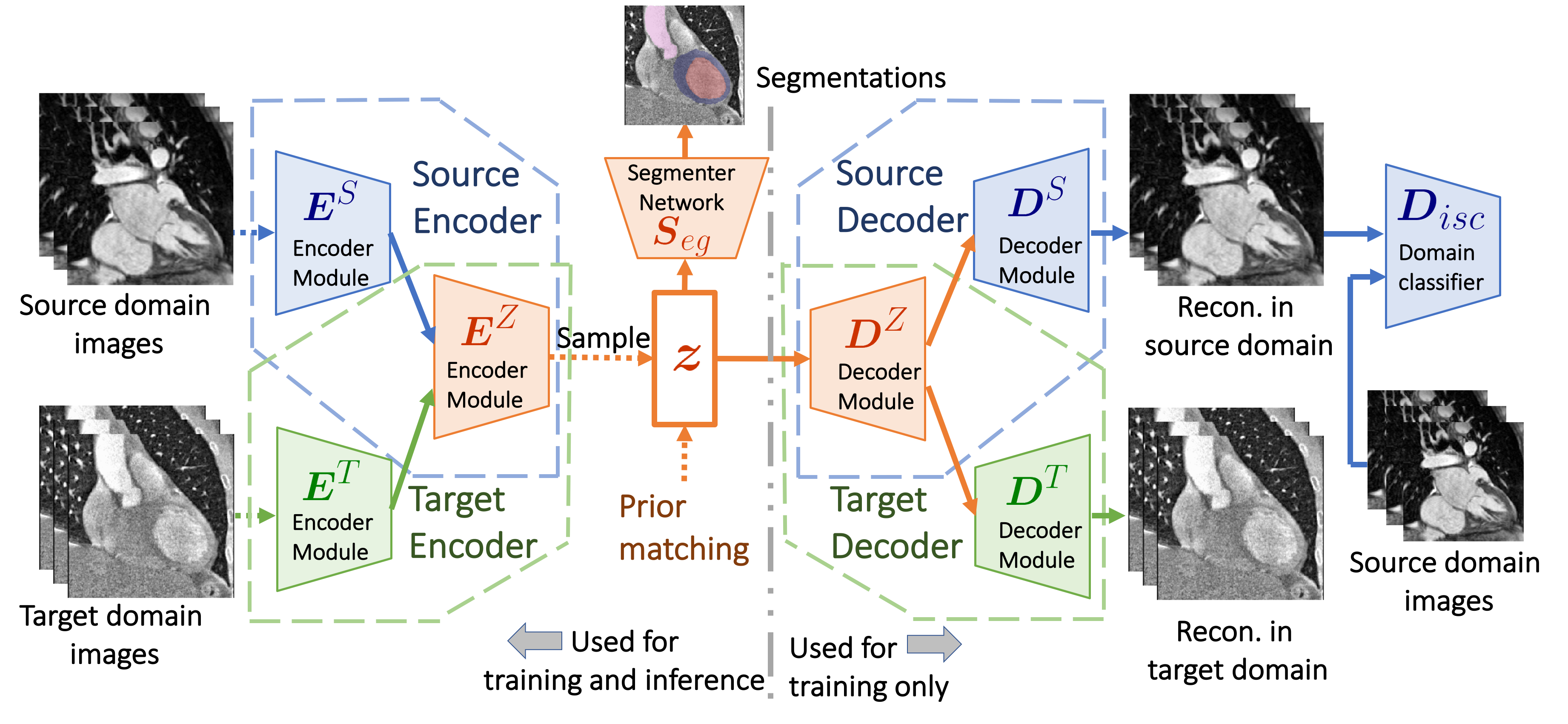}}
\caption{ Overview of the proposed architecture. A single image from the source domain or from the target domain is sent into its corresponding domain-specific encoder: $E^Z \circ E^S$ (source), or $E^Z \circ E^T$ (target). The encoder predicts the posterior of the latent feature $Z$ in $\mathcal{Z}$. The $S_{eg}$ then takes this as input. In training, we send the feature map to decoders $D^S \circ D^Z$ and $D^T \circ D^Z$ simultaneously to reconstruct images in both domains. The domain classifier network $D_{isc}$ then differentiates whether its input is from the original source image set or from the outputs of the source decoder.}
\label{fig:network}
\end{figure} 
\noindent\textbf{Overview:} The proposed method learns a feature space $\mathcal{Z} = \{ Z\}$, shared by both domains $\mathcal{D}_S$, $\mathcal{D}_T$, and the mapping $h(\cdot)$'s from input images $X$ from $\mathcal{D}_S$ or $\mathcal{D}_T$ such that $Z = h(X)$. It also learns a segmenter $S_{eg}$ from $\mathcal{Z}$ to label space $\mathcal{Y}$. For simplicity, we use subscripts $S$ and $T$ to refer to the domain of an image: $\mathit{e.g. }$ $\mx_S \sim P(X_S)$, of a mapping function: $\mathit{e.g.}$ $h_S(\cdot)$ and of a feature map sampled from its posterior in feature space: $\mathit{e.g.}$ $\mz_S \sim q(Z|\mx_S)$.

The overall architecture here consists of a VAE with two domain-specific encoders and decoders, which is extended from recent work by \cite{benaim2018one}, and a segmenter $S_{eg}$ operating on the VAE's latent space $\mathcal{Z}$ which will be learned to be domain invariant. The mappings $h(\cdot)$'s are realized with two encoders of the VAE. An overview of the network structure is illustrated in Fig. \ref{fig:network}. 
The posteriors $q(Z | \mx_S)$ or $q(Z | \mx_T)$ predicted by the source or the target encoder are modeled as multi-dimensional Gaussians $\mathcal{N}(\mu_Z, \Sigma_Z)$ with diagonal covariance matrices.
To train the model, $\mz \sim q(Z|\mx)$ is drawn at each iteration via the re-parameterization trick. It is then passed to both decoders to generate reconstructed images in two domains. 
Given an input $\mx_{S}$ from $\mathcal{D}_S$, we note as $\mx_{SS} = \mx_{S \rightarrow S}$ the reconstructed image in the same domain and as $\mx_{ST} = \mx_{S \rightarrow T}$ the output image in the other domain (\textit{vice versa} for input from $\mathcal{D}_T$). Meanwhile, $\mz$ is also used as input for the segmentation network $S_{eg}$.

\noindent\textbf{Supervised training in the source domain:} 
To obtain a source model as the basis for domain adaptation, we first train the VAE using the source encoder $h_S(\cdot) = E^Z \circ E^S: \mathcal{D}_S \rightarrow \mathcal{Z}$ and the corresponding decoder, in together with the segmenter network $S_{eg}: \mathcal{Z} \rightarrow \mathcal{Y}$ with source image-label pairs $\{ (\mx_S, \my_S) \}$. We have the VAE loss:
\begin{small}
\begin{align}
& \mathcal{L}^S_{vae} (E^S, E^Z, D^S, D^Z) =  \lambda_{rec} \, \mathcal{L}^S_{rec} + \lambda_{kl} \, \mathcal{L}^S_{kl} \nonumber \\
= & -\lambda_{rec} \, \mathbb{E}_{ \substack{ \mz_S \sim q(Z | \mx_S) \\ \mx_S \sim P(X_S) }  } [\log p (\mx_S | \mz_S) ] + \lambda_{kl} \, KL(q(\mz_S | \mx_S) || \mathcal{N}(0, I) ).
\label{equ: source_VAE}
\end{align}
\end{small}%\vspace{-0.17cm}

\noindent To overcome the class imbalance between relatively small segmentation labels and the large background, we employ a sum of soft Dice and weighted cross-entropy (CE) losses  to train $S_{eg}$ (which is common for medical image segmentation scenarios).
\begin{small}
\begin{align}
\mathcal{L}_{Seg}^S (S_{eg}, E^Z) = \mathbb{E}_{ \substack{ \mz_S \sim q(Z | \mx_S) \\ \mx_S \sim P(X_S) }  } [-Dice(S_{eg}(\mz_S), \my_S)  + CE(S_{eg}(\mz_S), \my_S)  ] .
\label{equ: source_seg}
\end{align}
\end{small}%\vspace{-0.17cm} 

\noindent In the meantime we pre-train the domain classifier $D_{isc}$ to classify whether its input is from the source training set $\{\mx_S\}$ or from reconstructed $\{ \mx_{SS}\}$ \cite{benaim2018one}. At present, $E^S$, $E^Z$, $D^Z$, $D^S$ are updated to minimize Eq.~ \ref{equ: adv_A} and $D_{isc}$ is updated to maximize Eq.~ \ref{equ: adv_A}. 
\begin{small}
\begin{align}
\mathcal{L}_{adv}^S (E^S, E^Z, D^Z, D^S, D_{isc}) = &  \mathbb{E}_{  \mx_S \sim P(X_S) } [ \log D_{isc}(\mx_S)]   + \nonumber \\ &  \mathbb{E}_{  \mz_S \sim q(Z_S | \mx_S)  }[ \log (1 - D_{isc}(  D^S(D^Z(\mz_S))  ) ) ].  
\label{equ: adv_A}
\end{align}
\end{small}

\noindent\textbf{UDA with prior matching:} 
The domain adaptation training starts after the source model is obtained. In addition to losses in Eq.~\ref{equ: source_VAE}-\ref{equ: adv_A}, we train the target encoding $h_T(\cdot) = E^Z \circ E^T$ and its decoding with a VAE loss. Similar to the process for the source domain, the posterior distribution $q(Z|\mx_T)$ in $\mathcal{Z}$ is predicted by feeding $\mx_T$'s to the target encoder. We therefore use the same form of VAE loss $\mathcal{L}^T_{vae}$ as that in $\mathcal{D}_S$ (Eq.~ \ref{equ: target_VAE}). To prevent $E^Z$ and $D^Z$ from overfitting on small $\{ \mx_T$\}, only $E^T$ and $D^T$ are updated \cite{benaim2018one}:
\begin{small}
\begin{align}
\label{equ: target_VAE}
\mathcal{L}^T_{vae} (E^T, D^T) = \lambda_{rec} \, \mathcal{L}^T_{rec} + \lambda_{kl} \, \mathcal{L}^T_{kl}.
\end{align}
\end{small}

\noindent We note that the regularizations $\mathcal{L}^S_{kl}$ and $\mathcal{L}^T_{kl}$ are particularly beneficial for data-efficient UDA. They match priors $P(Z_S)$ and $P(Z_T)$ by enforcing both priors to be close to $\mathcal{N}(0,I)$. We term this as \textit{prior matching} effect.

\noindent\textbf{UDA with domain adversarial training:} 
For domain adversarial training, we add $\mx_{TS}$'s into $D_{isc}$'s input set as fake examples.
\begin{small}
\begin{align}
\label{equ: adv_B}
\mathcal{L}_{adv}^T (E^T, D^T, D_{isc}) =   \mathbb{E}_{  \mx_S \sim P(X_S) } [ \log D_{isc}(\mx_S)] +  \mathbb{E}_{  \mx_{TS}  }[ \log (1 - D_{isc}(  \mx_{TS}  ) ) ], \nonumber \\
\text{where } \mx_{TS} = D^S(D^Z(\mz_T)), \, \mz_T \sim q(Z_T | \mx_T), \, \mx_T \sim P(X_T). 
\end{align}
\end{small}

\noindent To ensure two encoders providing aligned outputs for similar semantic information, we enforce cycle-consistency for images before and after encoding-decoding to a different domain \cite{zhu2017unpaired}. Unlike the common practice, cycle-consistency here is only applied on $\mathcal{D}_T \rightarrow \mathcal{D}_S \rightarrow \mathcal{D}_T$ direction, since mapping a large $\{\mx_S\}$ to a small $\{ \mx_T \}$ and mapping back is intuitively difficult in terms of preserving the large variety in visual appearance within $\{\mx_S\}$\cite{benaim2018one}. We therefore have:
\begin{small}
\begin{align}
\mathcal{L}_{cyc}^{T} (E^T, D^T) =  \mathbb{E}_{\mx_{T}}[ \Vert  D^T(D^Z( E^Z(E^S(\mx_{TS})))) - \mx_{T} \Vert_1 ], \nonumber\\
\text{where } \mx_{TS} = D^S(D^Z(\mz_T)), \, \mz_T \sim q(Z_T | \mx_T), \, \mx_T \sim P(X_T).
\label{equ: cyc_B}
\end{align}
\end{small}%\vspace{-0.17cm} 

\noindent The $D_{isc}$ therefore is trained to differentiate real source images $\{\mx_S\}$ against outputs of source specific decoder $\{ \mx_{SS}\} \cup \{\mx_{TS} \}$. 

We further propose to train $S_{eg}$ with $\{(\mz_{ST}, \my_S)\}$'s, where $\mz_{ST}$ is sampled from posterior obtained by sending $\mx_{ST}$ to the target specific encoder. This encourages $S_{eg}$ to be robust to remaining differences between $Z_T$ and $Z_S$ in $\mathcal{Z}$ due to imperfections of two encoders. We term this as \textit{task-consistency} as a straightforward analogy to cycle-consistency, which has also been independently found useful in \cite{hosseini2018augmented}.
\begin{small}
\begin{align}
\mathcal{L}_{cyc}^{task} (S_{eg}, E^Z) = \mathbb{E}_{\mz_{ST}} [-Dice(S_{eg}( \mz_{ST}) , \my_S ) + CE(S_{eg}(\mz_{ST}), \my_S)  ], \nonumber \\
\text{where} \,\ \mz_{ST} \sim q(Z | \mx_{ST}),\ \mx_{ST} = D^T(D^Z(E^Z(E^S(\mx_S)))),\ \mx_S \sim P(X_S).
\label{equ: cyc_task}
\end{align}
\end{small}%\vspace{-0.17cm} 

\noindent By summarizing Eq.~\ref{equ: source_VAE}-\ref{equ: cyc_task}, we have the entire training objective as follows:
\begin{small}
\begin{align}
\mathcal{L} = & \lambda_{rec} (\mathcal{L}^S_{rec} + \mathcal{L}^T_{rec})  + \lambda_{kl} ( \mathcal{L}^S_{kl} + \mathcal{L}^T_{kl}) + \lambda_{seg}( \mathcal{L}_{seg} +\mathcal{L}_{cyc}^{task}) \nonumber \\ & +\lambda_{adv}( \mathcal{L}^S_{adv} + \mathcal{L}^T_{adv}) + \lambda_{cyc} \mathcal{L}^T_{cyc}.
\label{equ: entire_loss}
\end{align}
\end{small}%\vspace{-0.17cm} 

\noindent\textbf{Model Implementation:} The network is implemented with PyTorch. $E^S$, $E^T$, $D^S$, $D^T$, $D^Z$ and $D_{isc}$ are configured as proposed in \cite{benaim2018one}. Although similar network structures have been used for unsupervised image translation \cite{benaim2018one,liu2017unsupervised}, the effects of various implementations of VAE on their tasks are often not studied in too much details. Unlike in some of popular implementations where the posterior covariance $\Sigma_Z$ is fixed to the identity matrix, in our implementation the last 2 blocks of $E^Z$ branch out to predict $\mu_Z$ and $\Sigma_Z$ maps separately. We have observed this design-of-choice yields the best performance by allowing the network to decide covariance $\Sigma_Z$ for different latent features. We employed a dilated residual network (DRN-26) \cite{yu2017dilated} for segmentation, with modifications on the front and the end layer configurations in adjust to our input and output sizes.
We simply chose the hyper-parameters of $\lambda_{rec}$, $\lambda_{cyc}$, $\lambda_{kl}$ and $\lambda_{adv}$ as 1.0, 10.0, 0.1 and 1.0 as proposed in \cite{benaim2018one}. We also refer readers to the work in \cite{liu2017unsupervised} on unsupervised image translation. Their network structure is similar to ours, and is shown to be relatively robust to different hyper-parameter selections. As $\mathcal{L}_{seg}$ and $\mathcal{L}^{task}_{cyc}$ are ordinary image segmentation losses, we simply set $\lambda_{seg}$ to be 1.0, same as common practices.
\section{Results}
\begin{figure}[]
\centering
\scalebox{0.97}{\includegraphics[width=1\textwidth]{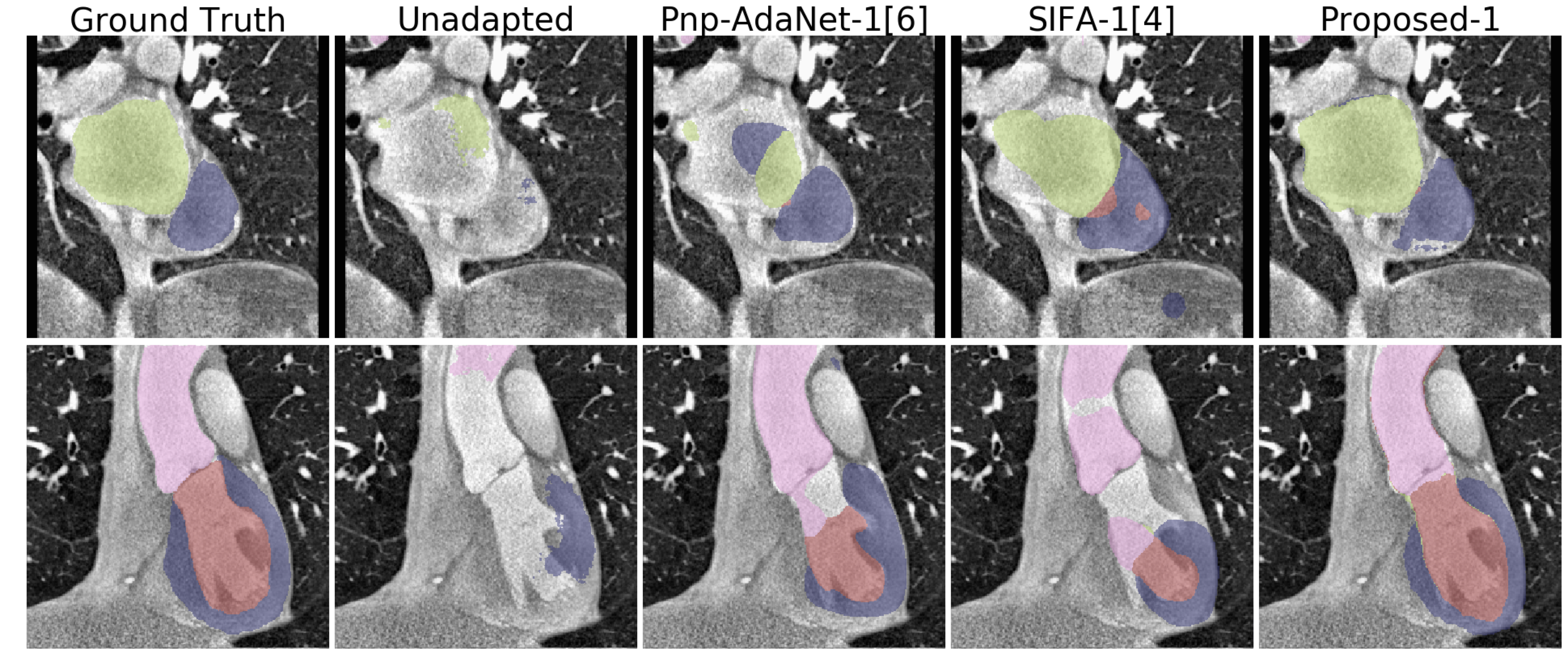}}
\caption{Qualitative results of adaptation performances on segmentation.}
\label{fig:seg_eval}
\end{figure}

\noindent\textbf{Dataset and training settings:} 
Our method is tested on the \textit{MICCAI Multi-modality Whole Heart Segmentation Challenge} \cite{zhuang2016multi} dataset. It contains 20 3D cardiac MRI and 20 3D CT scans from different clinical sites (note that that the MR and CT images are \textbf{unpaired}). Each CT contains $\sim$256 coronal slices while each MRI contains $\sim$128 after pre-processing. We chose the MR images as the source domain with sixteen labelled scans for training, four for testing whether the training on source domain functions properly. The CT images are taken as the target with sixteen scans chosen to create a pool for random selection for training while the remaining four scans are used for testing (as in \cite{chen2019synergistic,dou2018pnp}). Assignments from each individual scan to training or testing sets are kept the same as in \cite{chen2019synergistic,dou2018pnp}. Images are reformated as 2D along the coronal plane with a size of 3$\times$256$\times$256. Four cardiac structures including the left ventricle myocardium (LV-M), left atrium blood cavity (LA-B), left ventricle blood cavity (LV-B) and ascending aorta (A-A) constitute the segmentation labels. Small rotations, translations, shearings, elastic transformations, gamma transforms and intensity normalizations are used for data augmentation. The Dice score, and average symmetric surface distance (ASSD) of the largest 3D connected component for each label class, are employed for evaluation.  See Table \ref{tbl: eval} and Fig. \ref{fig:seg_eval} for quantitative and qualitative results.
\begin{table}[!t]
\centering
\caption{ Quantitative evaluations with the format $\genfrac{}{}{0pt}{1}{\text{mean}}{ \text{ (std.)} }$. Postfixes \textbf{-16} or \textbf{-1} after names of each method indicate the number of unlabelled target scans used for training. }
\scalebox{0.8}{
\begin{tabular}{ccccc|c|cccc|c}
\toprule
                            & \multicolumn{5}{c}{ \textbf{Dice} [\%] $\uparrow$}                    & \multicolumn{5}{c}{ \textbf{ASSD} [voxel] $\downarrow$}
                            \\ \cline{2-11} 
                            & LV-M    & LA-B    & LV-B    & A-A     & Mean   & LV-M    & LA-B    & LV-B    & A-A    & Mean \\ \hline 
\multirow{2}{*}{ \textbf{Oracle} }     & 82.35  & 88.45  & 89.28  & 87.92   & 87.00  & 6.03   & 8.26   & 7.08   & 1.61  &  5.74    \\
                            & (2.29) & (1.92) & (3.42) & (12.33) & (7.11) & (1.16) & (2.67) & (3.08) & (1.13)  &    (3.34) \\ 

\multirow{2}{*}{ \textbf{Unadapted}}  &  12.25      &  46.05      & 1.42       &    20.39     &     20.03   &  24.46      &     22.81   &   47.11     & 42.72       &  34.28   \\
                            &  (14.92)      &   (20.32)     &  (2.31)      &  (10.96)       &       (21.47) &    (12.07)  & (21.91)     & (17.91)       &  (12.61)     & (19.81)    \\ \hline       
\multirow{2}{*}{ \textbf{Pnp-AdaNet-16}\cite{dou2018pnp}} &    49.89     &  77.37      &    60.41     &    78.75     &    66.61   & 10.00      &    4.04    &     8.60   & 2.28  & 6.22  \\
                            &     (5.13)   &  (3.71)      &  (11.97)      &  (3.88)       &  (13.96)      & (3.20)      &  (0.76)      &  (1.93)      &  (0.84)      & (3.72)   \\
\multirow{2}{*}{ \textbf{SIFA-16}\cite{chen2019synergistic}}   &  63.58      &   80.03     &   79.90     & 79.58    &   75.77   &  3.44   &       3.89 &     3.31   &    2.64    & 3.32    \\
                            &   (3.30)     &     (3.21)   &   (7.51)     & (3.64) & (8.50)  &   (0.40)     &  (0.85)      &  (1.41)      &     (1.85)   &    (1.32)        \\    \hline                           
\multirow{2}{*}{ \textbf{Pnp-AdaNet-1}\cite{dou2018pnp}}       &    29.00    &     48.06   &  33.48      &     58.19    &    42.19    &    25.18    & 27.19       &    27.74    &      7.14  & 21.81     \\
                            &     (17.34)   &  (21.70)      &  (23.51)      &  (19.81)       &  (23.75)      & (28.10)      &  (37.18)      &  (28.70)      &  (5.43)      & (28.79)   \\                           
\multirow{2}{*}{ \textbf{SIFA-1}\cite{chen2019synergistic}}   &  39.76      &   76.65     &   53.36     & \textbf{80.27}    &   62.51   &  12.58   &       4.12 &     7.70   &    \textbf{2.72}    & 6.78    \\
                            &   (19.86)     &     (\textbf{6.29})   &   (24.42)     & (\textbf{5.81}) & (23.35)  &   (11.16)     &  (\textbf{1.19})      &  (4.45)      &     (\textbf{1.07})   &    (7.16)        \\
                            
\multirow{2}{*}{ \textbf{Proposed-1}}   &   \textbf{60.21}     &   \textbf{78.25}     &     \textbf{71.88}   &   78.38      &    \textbf{72.18}    &  \textbf{7.37}     & \textbf{3.87}      &    \textbf{6.44}    &  2.77      & \textbf{5.11}     \\
                            &   (\textbf{15.89})     &    (9.88)   &    (\textbf{17.93})    &  (13.21)        & (\textbf{16.31})        &      (\textbf{10.87})  &     (1.23)   &    (\textbf{2.18})    & (3.21)       & (\textbf{6.09})    \\  \bottomrule
\end{tabular}
}
\label{tbl: eval}
\end{table}

\noindent\textbf{Baseline and upperbound:} 
To illustrate the domain shift problem, we first obtained the 
\textbf{unadpated} baseline given by directly feeding target images to the source encoder after supervised training in the source domain. The result in Table \ref{tbl: eval} indicates that the source model completely failed on target images with a mean dice of 20.03\%. We also obtained the upperbound \textbf{oracle} by supervised fine-tuning on the source model with \textbf{all sixteen} target scans and their \textbf{labels}.

\noindent\textbf{Data-efficient domain adaptation:}
To simulate the scenario where only a small number of target data is avaliable, we here randomly draw only \textbf{one} scan from the target training pool and train the proposed data-efficient UDA. This is in drastic contrast to recent UDA works on this dataset, which use up \textbf{all sixteen} target scans for training \cite{chen2019synergistic,dou2018pnp}. To avoid being biased on one particular target training scan, the results shown on Table \ref{tbl: eval} are the averages of repeating the UDA training six times on different randomly chosen target scans. Compared with the unadapted baseline, a significant improvement by 52.15\% to 72.18\% in mean Dice is achieved. As shown In Fig. \ref{fig:seg_eval}, the proposed method yields results which are visually close to the ground truth. 

\noindent\textbf{Comparison with the state-of-the-art method:}
Under the same experiment setting and one-scan target training sample selections, we also compared our proposed method with two recent UDA methods which are specially designed for medical images: 
the \textit{Pnp-AdaNet}\footnote{\url{https://github.com/carrenD/Medical-Cross-Modality-Domain-Adaptation}} \cite{dou2018pnp} which is based on domain adversarial training \cite{tzeng2017adversarial}, and the recent state-of-the-art \textit{SIFA}\footnote{\url{https://github.com/cchen-cc/SIFA}} \cite{chen2019synergistic} which has been introduced in the \textit{Introduction} section. Table \ref{tbl: eval} shows that under the same target-data scarcity scenario, the proposed method in general outperforms the other two. We also include results of both methods trained on all sixteen target scans for reference. The proposed method obtains results which are close to those of \textit{SIFA-16}, but only require 1/16 of target data. 

\noindent\textbf{Ablation studies:} 
To highlight complementary effects of prior matching and domain adversarial training for UDA in faced of target data scarcity, we performed ablation studies by removing each of these two components separately. By removing one of prior regularization or domain adversarial training, the model easily overfit. The performances measured by mean Dice drop to lower than 55\% and they oscillate instead of converge. 

\noindent\textbf{Toward few-shot UDA:}
%Few-shot learning usually refers to as training with a few ($\mathit{e.g. }$ $<$10) different samples.
We experimented adaptation with a few target \textbf{2D slices} by training with only 3 consecutive target slices intersecting with three of four labels. The model overfits eventually. Nevertheless, by applying early stopping after tens of epochs, it could still realize a mean Dice of over 60\%.
\section{Conclusion and Discussion}
We present a novel data efficient unsupervised domain adaptation method for medical image segmentation which overall outperforms the state-of-the-art method given only a small target set. Unlike most of previous UDA methods which use plain encoder-decoder networks and focus on pure domain adversarial training, we demonstrate the effectiveness of VAE-based prior matching in faced of target data scarcity. Although not upperbounding them (KL-divergence does not satisfy the triangular inequality), 
$\mathcal{L}_{kl}^S + \mathcal{L}_{kl}^T$ provides approximations of KL-divergences between prior distributions of features from two domains, which are in principle extremely difficult to directly estimate given the small target set. By independently forcing source and target prior distributions to be close to $\mathcal{N}(0, I)$, we can match distributions between domains better. From the perspective of data augmentation, sampling from posteriors with noise augments the data. This naturally improves model robustness by introducing further variabilites and perturbations when training its downstream network components. 

Our UDA work also differs from VAE-based unsupervised image translation (UIT)\cite{benaim2018one,liu2017unsupervised}. Instead of obtaining high-quality transferred images or disentangled representations for manipulation as in UIT, our method focuses on obtaining distribution-matched latent semantic features, and therefore manages to fully exploits the prior matching effect described above. 
Finally, we performed an extreme test on few-shot UDA with the hope to inspire future studies.
\subsubsection{Acknowledgement}
This work is supported by the EPSRC Programme Grant (EP/P001009/1).
\bibliographystyle{splncs04}
\bibliography{paper589.bbl}
\end{document}